\documentclass[12pt]{JHEP3}

\usepackage{epsfig}
\epsfclipon
\usepackage{multicol}
\usepackage{epsfig,bm}
\usepackage{amssymb,amsmath}

\newbox\mybox
\newcommand\fverb{\setbox\mybox=\hbox\bgroup\verb}
\newcommand\fverbdo{\egroup\medskip\noindent\fbox{\unhbox\mybox}\ }
\newcommand\fverbit{\egroup\item[\fbox{\unhbox\mybox}]}

\newcommand{\nco}{\newcommand}
\nco{\beq}{\begin{equation}} \nco{\eeq}{\end{equation}}
\nco{\beqa}{\begin{eqnarray}} \nco{\eeqa}{\end{eqnarray}}
\nco{\lra}{\leftrightarrow}

\nco{\sss}{\scriptscriptstyle} \nco{\dphi}{\varphi}
\nco{\lsim}{\mbox{\raisebox{-.6ex}{~$\stackrel{<}{\sim}$~}}}
\nco{\gsim}{\mbox{\raisebox{-.6ex}{~$\stackrel{>}{\sim}$~}}}

\def\pref#1{(\ref{#1})}

\def\pref#1{(\ref{#1})}

\title{Supersymmetric Large Extra Dimensions and the Cosmological
Constant: An Update}

\author{C.P.~Burgess\\

        Physics Department, McGill University,\\
        3600 University Street, Montr\'eal, Qu\'ebec, Canada H3A
        2T8.}

\abstract{This article critically reviews the proposal for
addressing the cosmological constant problem within the framework
of supersymmetric large extra dimensions (SLED), as recently
proposed in {\tt hep-th/0304256}. After a brief restatement of the
cosmological constant problem, a short summary of the proposed
mechanism is given. The emphasis is on the perspective of the
low-energy effective theory in order to see how it addresses the
problem of why low-energy particles like the electron do not
contribute too large a vacuum energy. This is followed by a
discussion of the main objections, which are grouped into the
following five topics: (1) Weinberg's No-Go Theorem. (2) Are
hidden tunings of the theory required, and are these stable under
renormalization? (3) Why should the mechanism apply only now and
not rule out possible earlier epochs of inflationary dynamics? (4)
How big are quantum effects, and which are the most dangerous? and
(5) Even if successful, can the mechanism be consistent with
cosmological or current observational constraints? It is argued
that there are plausible reasons why the mechanism can thread the
potential objections, but that a definitive proof that it does
depends on addressing well-defined technical points. These points
include identifying what fixes the size of the extra dimensions,
checking how topological obstructions renormalize and performing
specific calculations of quantum corrections. More detailed
studies of these issues, which are well reach within our present
understanding of extra-dimensional theories, are currently
underway. As such, the jury remains out concerning the proposal,
although the prospects for acquittal still seem good. (An abridged
version of this article appears in the proceedings of {\it SUSY
2003}.)}

\begin{document}

\section{What is So Hard About the Cosmological Constant Problem?}

The current evidence for a small but nonzero cosmological constant
\cite{ccnonzero} underlines the complete absence of a convincing
theoretical understanding for why this constant of nature can be
so small. On the one hand, observations now indicate the vacuum
energy density is of order $\rho = v^4$, with $v \sim 3 \times
10^{-12}$ GeV in units for which $\hbar = c = 1$. On the other
hand all known theories of microscopic physics appear to predict
that a particle of mass $m$ contributes to $\rho$ an amount which
is of order $\delta \rho \sim m^4$. The problem arises because
almost all currently known elementary particles have masses much
larger than $10^{-12}$ GeV. (See \cite{ccreview} for a review of
past heroic attempts to solve this problem, together with a no-go
theorem.)

As is the case for any `naturalness problem', any fundamental
solution to the cosmological constant problem must answer two
questions.
\begin{enumerate}
\item
Why is the vacuum energy, $\rho_0$, so small at the microscopic
scales, $M \gsim 10^3 \, \hbox{GeV}$, at which the fundamental
theory is couched?
\item Why does it remain small as all the scales
between $M$ and $v$ are integrated out?
\end{enumerate}
It is the second of these problems which is the more worrying,
because it seems to indicate that we are missing something in our
description of {\it low energy} physics, which we normally think
we understand. For instance, we believe we understand the electron
quite well and yet even in this supposedly well-understood theory
we do not know why the electron doesn't contribute the
catastrophically large amount $\delta \rho \sim m_e^4$, with $m_e
= 5 \times 10^{-4}$ GeV. At present our only theoretical refuge is
that of the desperate: we claim that the microscopic theory
predicts an enormous cosmological constant, $\rho_0 \sim M^4$, and
that this gets systematically cancelled (to at least 60 decimal
places!) by the process of integrating out all of the lighter
particles down to the very low energies where $\rho$ is measured.

Supersymmetric theories provide the only partial ray of light in
this otherwise totally dark picture. This is because sufficiently
many supersymmetries can explain why the microscopic value,
$\rho_0$, must be small or vanish (thereby addressing problem 1).
It can also partially explain why the process of integrating out
lighter particles does not ruin this prediction (problem 2),
because supersymmetry enforces a cancellation between bosons and
fermions in their contributions to $\rho$. Unfortunately, this
cancellation is only partial if supersymmetry is broken, leaving a
residual value which can be as small as $\delta \rho \sim
m_{sb}^4$, where $m_{sb}$ is a measure of the largest mass
splittings between bosons and fermions within a supermultiplet.
Sadly, experiment already implies that this scale must be at least
as large as $m_{sb} \sim 10^{2}$ GeV for observed particles like
electrons.

\section{Supersymmetric Large Extra Dimensions (SLED)}

Problems like the cosmological constant problem, which resist
solution for many years, normally do so because they involve a
hidden assumption which everyone has made, and which unnecessarily
restricts the kinds of theories which could be considered. If only
such an assumption could be identified, one could ask whether its
relaxation might help reduce the size of the predicted
cosmological constant.

Crucially, any such assumption must already be playing a role in
the description of physics above the scale $v \sim 10^{-12}$ GeV,
set by the cosmological constant itself. At present this
requirement is a hindrance because physics at these scales is
heavily constrained by observations, making the search for the
successful theory more difficult. Ultimately it will be a boon,
however, since any successful proposal must change physics at
these low energies in many ways, and so must inevitably make
testable low-energy predictions in addition to explaining why
$\rho$ is so small. This is certainly true of the scenario which
is of interest in this paper.

\subsection{The Assumption of 4 Dimensions}

In ref.~\cite{branesphere} it is proposed that the hidden
assumption which might be behind the cosmological constant problem
is the assumption that physics is four dimensional at energy
scales above $v$. After all, this assumption is crucial to the
problematic statement that a particle of mass $m$ contributes an
amount $\delta \rho \sim m^4$. If the world were to involve extra
dimensions as large as $r \sim 1/v \sim 0.1$ mm, then any
calculation of the contribution of a particle having mass $m > v$
must be done in a higher-dimensional theory for which the
predictions may differ. This supersymmetric
large-extra-dimensional (SLED) picture is based on the earlier
non-supersymmetric large-extra-dimensions (LED) proposal
\cite{add}, the authors of which first realized that the world
could be extra-dimensional down to such extremely low energies.

Extra dimensions can be as large as $r \sim 1/v$ without being in
conflict with experimental observations provided that one works
within the string-motivated brane-world framework, where all
observed interactions except gravity are trapped on a
4-dimensional surface (brane) within a 6-dimensional spacetime
\cite{add}. Since in this picture only gravitational interactions
can probe the existence of the extra two dimensions, these
dimensions can be much larger than had been hitherto believed. It
happens that the current limit on the size of these two
dimensions, based on the best present measurements of the
gravitational inverse-square law \cite{gravbounds}, is about $r
\sim 1/v \sim 0.1$ mm.

The SLED picture can work, but only if there are precisely two
extra dimensions which are as large as $1/v$, and only if the
scale of gravitational physics in the extra dimensions is as low
as is possible: $M \sim 10^3$ GeV. This last requirement is
required because Newton's constant, $8 \pi G = 1/M_p^2$, is
related to $r$ by $M_p \sim M^2 \, r$, and so $M = (M_p /r)^{1/2}
\sim 10^3$ GeV.\footnote{We use here the Jordan frame, for which
$M_p$ is $r$ dependent but the electroweak scale, $M$, is not. It
is in this frame that the Kaluza-Klein mass scale is $m_{\rm KK}
\sim 1/r$.} Since $M$ is near the weak scale, it follows that any
SLED proposal is likely to have an interesting phenomenological
signature in collider experiments at the LHC
\cite{susyaddbounds,susyaddbounds2} just as is true for the LED
picture \cite{PDG,realgraviton,virtualgraviton}.

Both the LED and SLED scenarios are also subject to astrophysical
bounds \cite{PDG,LEDastrobounds,LEDcosmobounds,susyaddbounds}, but
we temporarily put these aside to see whether progress can be made
with the cosmological constant. If so, we are confident that
model-dependence of the astrophysical arguments can be exploited
to evade the astrophysical bounds.\footnote{See, for instance,
ref.~\cite{rabi} for a sample way around some of these
constraints.}

\subsection{Classical Contributions to $\rho$}

Can extra dimensions help us understand why the electron does not
contribute too large a vacuum energy density? It turns out that
they can. Since observations require that supersymmetry is badly
broken on the brane, within the SLED framework integrating out the
electron (or any other observed particles) indeed does contribute
a vacuum energy density which is of order $m^4$. But this energy
density is {\it not} a cosmological constant. Rather, it is a
contribution to the tension of the brane: $\delta T \sim m^4$.
That is, it is an energy source which is localized within the
extra dimensions at the position of the brane on which we live. We
must ask how this energy source curves the extra dimensions, and
then see what the implications are for the effective 4D
cosmological constant which would be observed on distance scales
larger than $r \sim 0.1$ mm.

It happens that the curvature of the extra dimensions due to this
localized energy source can be computed, with the result that the
geometry acquires a conical singularity at the position of the
branes, with a correspondingly singular contribution to the
two-dimensional curvature given by\footnote{This and later
expressions use Weinberg's curvature conventions \cite{GandC}}
\beq \label{Ricci2}
    R_2 = - {2\over e_2} \, \sum_i T_i \,
    \delta^2(y-y_i) + \dots \, .
\eeq
Here $e_2 = \sqrt{\, \det g_{mn}}$ is the volume element for the
internal two dimensions, $y_i$ denotes the position of the `$i$'th
brane and the ellipses denote contributions to $R_2$ which are
smooth at the position of the brane.

How do the tension and curvature contribute to the effective 4D
cosmological constant on long distance scales? This is obtained by
integrating out the bulk gravitational degrees of freedom, which
are not localized on the branes. If these are integrated out at
the classical level an interesting cancellation occurs between the
brane tensions and the bulk curvature. The effective 4D
cosmological constant obtained at this order is
\beqa \label{rhocl}
    \rho_{\rm cl} &=&  \sum_i T_i + \int_M d^2y \; e_2 \,
    \left[\frac12 \, R_2 + \dots \right] \nonumber \\
    &=& 0 \, ,
\eeqa
where the sum on `$i$' is over the various branes in the two extra
dimensions and `$\dots$' denotes all of the other terms besides
the Einstein-Hilbert term in the supersymmetric bulk action.
Interestingly the sum over brane tensions, $T_i$, precisely
cancels the contribution of the singular part of the curvature,
eq.~\pref{Ricci2}, to which they give rise \cite{clp}. Remarkably,
the same kind of cancellation also occurs amongst the remaining
terms in $\rho_{\rm cl}$ once these are evaluated using the smooth
parts of the geometry and the other bulk fields obtained using the
classical field equations \cite{branesphere}.

We see there is a cancellation in $\rho$ between the contribution
of the brane tensions and the extra-dimensional curvatures to
which these tensions give rise. Better yet, this cancellation does
not depend at all on the {\it value} of the brane tension, and so
applies equally well even if the $T_i$ are large and include all
of the quantum effects due to virtual particles localized on the
branes.

\subsection{Bulk Loops}

Given that classical contributions cancel, what of the quantum
effects? Since the quantum effects on the brane can be regarded as
being included in the value of the effective brane tension, the
quantum effects which remain are those involving the particles
which live in the bulk. These include the graviton and all of its
partners under higher-dimensional supersymmetry.

These bulk particles have several properties which are crucial for
understanding the size of their contributions to the effective 4D
cosmological constant, $\rho$, and which follow from the fact that
they are related by supersymmetry to the graviton. The most
important such property is the scale of supersymmetry breaking in
the bulk, since this governs the effectiveness of the quantum
cancellations in the effective 4D cosmological constant.
Remarkably, this scale is naturally of order $m_{sb} \sim v$ as
may be seen by either of two arguments. First, this scale follows
because supersymmetry breaks on the brane at scale $M$, to which
each bulk mode only couples with (4D) gravitational strength
(because they are related to the graviton by supersymmetry).
Consequently $m_{sb} \sim M^2/M_p \sim v$. Alternatively it
follows from an explicit Kaluza-Klein reduction, because there is
one massless graviton mode but the presence of the branes forces
the lightest gravitino mode to be of order the Kaluza-Klein scale,
$m_{sb} \sim 1/r \sim v$. The agreement of these two lines of
reasoning relies on the value of $r$ required by the hierarchy
problem, inasmuch as this ensures $M^2/M_p \sim 1/r$.

The SLED proposal for understanding the small cosmological
constant is founded on the assertion that the quantum
contributions of bulk fields to the vacuum energy are of order
$\delta \rho \sim m_{sb}^4$. In this case we see that bulk loops
would predict a cosmological constant whose magnitude agrees with
observations. Although $m_{sb}^4$ is arguably a reasonably generic
estimate for a vacuum energy in a supersymmetric theory, a key
part of the argument is a detailed justification of this estimate.
This is particularly so, given that the supersymmetry-breaking
fields are extra-dimensional particles which do couple
nontrivially to the supersymmetry-breaking physics on the branes.
A closer examination of this point is one of the discussion items
in the next section.

There is indirect evidence that the simple estimate $\rho \sim
m_{sb}^4$ is correct. This evidence comes from direct one-loop
string-based calculations of the vacuum energy in toroidal
compactifications with supersymmetry broken by boundary conditions
and/or the presence of branes \cite{ablm}. Such explicit
calculations indeed find a vacuum energy which is of order $\rho
\sim m_{sb}^4 \sim 1/r^4$. These calculations leave several issues
open, however, such as whether the result is an artefact of the
one-loop approximation or of the use of toroidal geometry. It
would be preferable to be able to complement them with more
general arguments based on power-counting within the low-energy
theory, and so we return to this issue in the next section.

\section{A Closer Inspection}

Although the picture presented in the previous section is
appealing, it must address several issues in order to be accepted
as a real solution to the cosmological constant problem. These
issues are summarized in this section, together with a discussion
of the likelihood that they are likely to be addressed by the SLED
proposal.

\subsection{Scale Invariance and Weinberg's No Go Theorem}

The most interesting feature about the classical vanishing of
$\rho$ described above is that the cancellation between brane and
bulk contributions occurs for {\it any} value of brane tension.
This is reminiscent of a similar cancellation which also has been
found to occur in some 5-dimensional models \cite{5dst}, and one
wonders whether they occur for similar reasons. It turns out they
are indeed similar, because in both cases the cancellation of the
effective 4D cosmological constant can be traced to the existence
of an underlying classical scale invariance of the bulk gravity or
supergravity action. Provided this symmetry is not broken by the
brane couplings, the classical contribution to $\rho$ in 4D is
guaranteed to vanish \cite{susyads}. (More precisely, for brane
actions which --- in the 6D Einstein frame --- take the form
\beq \label{lambdaeq}
    S_b = - T_3 \int d^4x \; e^{-\lambda \phi} \, \sqrt{-g} \,,
\eeq
where $\phi$ is the dilaton which appears in the 6D gravity
supermultiplet, the condition that the brane couplings do not
break the classical scale invariance of the bulk theory requires
$\lambda = 0$.)

Because the cancellation leading to vanishing $\rho$ in the
classical integration over bulk fields relies on a classical scale
invariance, the cancellation need not survive the quantum part of
the bulk integration. The difference between SLED and the 5D
models is that for SLED there is a bulk supersymmetry which can
plausibly protect the 4D cosmological constant from receiving
corrections which are too large, since the supersymmetry breaking
scale in the bulk is precisely $m_{sb} \sim v$. There is no such
symmetry at work in the 5D models (or models in other dimensions),
and so for these theories there is no reason to expect that the
self-tuning should survive quantum effects in the bulk.

The reliance on classical scale invariance also brings to mind
Weinberg's No-Go theorem. In ref.~\cite{ccreview} Weinberg raises
a very general objection against so-called self-tuning solutions
to the cosmological constant problem, which in their essence rely
on scale invariance of the underlying equations. In cartoon form,
his objection proceeds as follows. Even if scale invariance is a
good symmetry at the quantum level, it must be spontaneously
broken because physical scales like particle masses are known to
exist. There must therefore be a goldstone (or pseudo-goldstone)
boson for scale invariance, $\sigma$,\footnote{Confusingly,
$\sigma$ is called a dilaton, but should not be confused with the
6D scalar field, $\phi$, which is related to the metric by
supersymmetry in six dimensions.} which transforms under scale
transformations like $\sigma \to \sigma + \epsilon$. But although
this transformation can forbid a cosmological constant, it does
not solve the cosmological constant problem since it cannot
prevent the generation of a scalar potential of the form $V_{\rm
dil} \propto e^{c\sigma}$, for some constant $c$. However if such
a potential exists, then $\sigma$ gets driven to infinitely large
values, which corresponds to the scale-invariant vacuum ({\it
i.e.} the one which does not spontaneously break scale
invariance). Although such a theory does have a vanishing
cosmological constant, it does so at the cost of having all masses
vanish because of the unbroken scale symmetry. This is not a
successful theory of the world which is revealed to us by
experiments.

Because the argument is explicitly couched in 4 dimensions, it
cannot describe the generation of a localized brane tension by
integrating out brane particles (like the electron above). It can
and does describe the physics of the cancellation between the
resulting localized tension on the branes and the curvature which
these tensions set up in the extra dimensions. It does so because
this classical cancellation is based on a classical scale
invariance \cite{susyads}. Weinberg's argument then correctly says
that this invariance cannot save us from quantum corrections which
generate a potential energy for the dilaton, $\sigma$. This is
true, and in the SLED case implies that the quantum corrections
are likely to generate a potential for the classical flat
direction which is parameterized by the particular combination of
the radius and 6D dilaton field, $s = e^\sigma = e^\phi/r^2$
\cite{susysphere,gp}. Understanding this potential is a crucial
step towards understanding what dynamics stabilizes the size of
the extra dimensions to $r \sim 1/v$. Because it is supersymmetry
-- and not scale invariance -- which in the SLED scenario enforces
the bulk bose-fermi cancellations which keep $\rho$ small, this
part of the mechanism is not affected by Weinberg's argument.
Furthermore, because the scale invariance is broken by quantum
effects, the potential which bulk loops produce need not have the
simple form $e^{c\sigma}$, and so need not necessarily imply a
runaway to a scale invariant vacuum. (Whether it actually does
predict such a runaway is a more detailed question, more about
which in later sections.)

\subsection{Hidden Fine Tunings?}

An advantage of the 6D SLED proposal is the ability to solve the
back-reaction problem and write down explicit solutions to the
bulk equations of motion, including the gravitational effects of
the branes. As such it is possible to more closely examine the
nature of the solutions, and so to see if the classically
vanishing vacuum energy has somehow been ensured through a hidden
fine-tuning of some of the parameters of the theory. This kind of
hidden fine-tuning has been been raised as an objection
\cite{5dtune} to the 5D models of ref.~\cite{5dst}.

Indeed, at first sight there does seem to be such a tuning going
on in the SLED solutions \cite{branesphere,navarro}. It is the
purpose of this section to describe this tuning, and to argue why
it need not be a problem for the SLED proposal for the
cosmological constant problem.

For these purposes a concrete example is helpful, and the simplest
one to think about in this regard is the `rugby ball' solution of
the gravity \cite{rugbygrav} or supergravity \cite{branesphere}
equations. Here the geometry of the internal dimensions is a
2-sphere threaded by the flux of a magnetic monopole, which
generalizes an old compactification \cite{ss,rbd} of 6D
Nishino-Sezgin supergravity \cite{ns}. The solution includes two
branes, which can be located at the sphere's north and south pole.
Their back-reaction onto the geometry is then described by
removing a wedge from the two sphere and identifying opposite
sides of the wedge. Einstein's equations imply that the angular
width of the defect at each end of the wedge must be strictly
proportional to the tension of the brane at that end. But such a
wedge can only be removed from a sphere without curving it if the
boundaries of the wedge are lines of fixed longitude on the
sphere, and so the defect angle at both ends of the wedge must in
this case be equal. The spherical solution therefore requires the
tension of the two branes to be positive and to be precisely
equal.

The bulk geometry can also be found for 6D supergravity even if
the brane tensions are not equal \cite{ggp}. It is found in this
case that both the warping of the 4D metric and the 6D dilaton
field acquire a nontrivial variation across the extra dimensions.
There is a generalization of the equal-tension constraint also for
this geometry, however, which relates the tensions of the two
branes to various bulk quantities. This constraint simply
expresses the topological statement that the Euler number of the
internal geometry is the same as for a 2-sphere
\cite{ggp,susyads}.

A similar constraint may also be derived involving the gauge
coupling constants and the amount of magnetic flux threading the
sphere \cite{branesphere}, and this second constraint also has a
topological interpretation. It expresses the statement that the
Chern class of the 2-dimensional magnetic field is the same as for
a magnetic monopole.

Relations such as these between the tensions of the branes are
also reminiscent of the situation in 5 dimensions. There,
solutions are found to the bulk Einstein equations in the presence
of a single brane, for which the effective 4D geometry is also
flat regardless of the value of the brane tension. However in this
case the bulk solutions are found necessarily to be singular, and
this singularity can be interpreted as the response of the
geometry to a second brane whose existence is required by the
solution. Since it turns out that the tension of this second brane
is equal and opposite to the tension of the first brane, their
tensions cancel in the 4D vacuum energy and the flatness of the 4D
space is explained by an apparent self-tuning. This 5D constraint
of opposite tensions for the branes is also topological in nature.

Does the existence of such constraints kill the SLED proposal for
solving the cosmological constant problem? Not necessarily, for
the following reasons.\footnote{There is a {\it caveat} to the
argument given here, concerning the warping of the bulk geometry,
which is discussed in a later section.}

The central point to keep in mind here is version 2 of the
cosmological constant problem as given above. One issue which this
raises is whether the effective 4D cosmological constant remains
zero as successive scales are integrated out from a high scale,
say $M \sim 10^3$ GeV to the low scale $v \sim 10^{-12}$ GeV. We
must ask: If a constraint among the tensions or magnetic fluxes is
imposed in the short distance theory at distances of order $1/M$,
does it remain imposed as we renormalize down to long distances of
order $1/v$? If so, then an understanding of the smallness of the
cosmological constant can be consistently deferred until the
theory describing physics at energy $M$ is understood.

Although this renormalization has not yet been performed
explicitly for these models, there is a simple line of reasoning
which argues that the constraints we are considering (like
equality of the tensions) should be stable against integrating out
the scales between $M$ and $1/r$. The key point is that all of
these constraints are topological in origin, in the 6D case being
related to the expression for the Euler number or the Chern class
of the internal field configurations. They take the generic form
$F(T_1,T_2,r,...) = n$, where the left-hand-side is a function,
$F$, of the brane tensions and charges as well as of various bulk
quantities. Because the right-hand-side is an integer, the
quantity $F$ cannot renormalize even if its arguments do. As such,
these constraints cannot be changed by the integrating out of
modes whose wavelengths are very short compared with the size $r$
of the internal space. Integrating out very short wavelength modes
can renormalize {\it local} operators in the 6 dimensions, but do
not `know' about the topology because their wavelengths do not
reach completely across the internal space.\footnote{It is this
property which underlies the general expression of ultraviolet
effects in terms of a local curvature expansion \cite{Gilkey}.}

There is an indirect check that the vanishing of the classical
vacuum energy is stable to perturbations to special features of
specific solutions (like equality of tensions on the two branes).
The check comes from the discovery of a very general class of
axially-symmetric solution for the two-brane configuration in 6D
supergravity \cite{ggp}. These describe the back-reaction which is
appropriate to an arbitrary pair of brane tensions (but excludes
the types of direct dilaton and magnetic couplings to the branes
discussed in ref.~\cite{branesphere}), and so allow one to ask
whether the classical tension-cancellation mechanism remains
satisfied in a wider context. Most remarkably, despite their not
being supersymmetric, for {\it all} of these solutions the 4
dimensions seen by brane observers are found to be precisely flat,
regardless of the relative size of the tensions on the two branes,
in agreement with the general scale invariance arguments of
ref.~\cite{susyads}.

\subsection{The `Moving Target' Problem}

Is the SLED mechanism too much of a good thing? That is, does it
act to zero the effective cosmological constant at all times
during cosmology or just at present? If it were to zero the
cosmological constant at all times it would be difficult to
understand how inflation could ever arise in the universe's
history -- what is being called here the `moving target' problem.

That this is not a worry follows from the observation that the
radius, $r$, of the extra dimensions is unlikely to always have
been as large as it is now. Although $r$ cannot have changed much
since the epoch of Big Bang Nucleosynthesis (more about this
later), it is very likely to have evolved considerably ---
probably growing from much smaller values --- at much earlier
times. If so then the prediction $\rho \sim 1/r^4$ need not imply
that $\rho$ must be negligibly small at epochs of cosmology
earlier than nucleosynthesis.

A related question asks what happens if the tensions on the two
branes in the rugby ball solution suddenly change, as they would
once one considers them to be field-dependent quantities
associated with the scalar potential for various fields localized
on the branes. In particular, there could be vacuum phase
transitions taking place on the branes, such as the electroweak or
QCD phase transition, or an earlier transition possibly associated
with baryogenesis or other higher-energy physics. Since the
tensions on the two branes are unlikely to evolve identically, how
can the various topological constraints continue to be satisfied?

It is possible to answer this issue in some detail at the
classical level, because of the discovery of the more general
geometries containing a pair of branes having different tensions
\cite{ggp,susyads}. If the brane tensions were to spontaneously
change then the bulk geometry is forced by the topological
constraints to warp in response. Although this warping does not
change the classical cancellation of the brane tensions in the
effective 4D cosmological constant, it likely {\it does} ruin the
argument that the quantum contributions remain as small as
$10^{-12}$ GeV. This is because in the warped case the basic
estimate for the quantum corrections remains $\delta \rho \sim
m^4_{\rm KK}$, where $m_{\rm KK}$ is a typical Kaluza-Klein mass
for the bulk modes. Now, in the unwarped case this mass must be
very small, $m_{\rm KK} \sim v \sim 1/r$, in order to have the
correct hierarchy between the strength of the electroweak and
gravitational interactions ({\it i.e.}~solution to the hierarchy
problem). The KK mass need {\it not} be this small for the warped
geometries, since for these part of the electroweak hierarchy can
be understood in terms of the warping itself, {\it \`a la} Randall
and Sundrum \cite{rs}. As a result, for warped geometries the size
of the internal dimensions which is required to ensure the correct
electroweak hierarchy implies a Kaluza-Klein scale $m_{\rm KK}$,
which is much {\it larger} than $v$. As such, the naive estimate
$\delta \rho \sim m^4_{\rm KK}$ is also much larger for these
geometries than the presently-observed dark energy density.

This shows that if the SLED is to provide a successful explanation
of the presently-small cosmological constant, it must explain why
at present the extra-dimensional geometry is large enough to
explain the electroweak hierarchy {\it and} why the internal
dimensions are not warped. This is a dynamical issue whose
resolution requires an understanding of the as-yet-unsolved issue
of radius stabilization in these models. (See, however,
ref.~\cite{Hall,ABRS1} for some first steps towards understanding
this issue within the SLED context.)

\subsection{How Big Are Quantum Effects?}

The central issue in the SLED proposal is to justify the estimate
$\delta \rho \sim m_{sb}^4$ of the contribution to $\rho$ due to
quantum effects involving loops of bulk fields. Although this is
most properly done through explicit calculations using geometries
such as the rugby ball, along the lines of ref.~\cite{ablm}, there
is much to be learned from explicit power-counting arguments.

As described earlier, the generic 4D estimate that integrating out
a particle of mass $M$ generates a vacuum energy density of order
$M^4$ {\it does} apply to the energy density generated by
integrating over the 4D brane-bound modes. This naturally leads to
brane tensions which are of order $M^4$, where $M$ is a scale in
the TeV region. Crucially, in SLED this is a localized energy
distribution which sets up an extra-dimensional gravitational
field, as opposed to being a direct contribution to the effective
4D cosmological constant. As such it is cancelled by the classical
integration over the bulk modes, as argued above and in
ref.~\cite{branesphere}.

At first sight the quantum integration over bulk modes is more
problematic, since these can be regarded as an infinite
Kaluza-Klein tower of 4D modes. Although some of these modes are
comparatively light --- with masses of order $m_{sb} \sim v$ ---
there are elements of the tower which are as much more massive,
including masses of order $M$. One might expect the sum over all
such modes to again contribute a contribution of order $M^4$. Of
course additional symmetries like supersymmetry can ameliorate
this conclusion by enforcing cancellations between bosons and
fermions, and this requires $\delta \rho$ to vanish in the limit
that the bulk supersymmetry-breaking scale, $m_{sb}$, goes to
zero. But in itself this could allow contributions of order
$\delta \rho \sim M^2 m_{sb}^2$, and although these are much
smaller than $M^4$ they are much larger than $m_{sb}^4$.

A more geometrical idea of how scales like $M^2 m_{sb}^2$ might
arise can be had by considering the types of local effective
interactions which can be generated by integrating out a bulk mode
whose mass is of order $M$. The key difference for SLED over other
cosmological constant proposals is that 6D general covariance
requires this effective interaction to be local in six-dimensions,
for all $M$'s right down to the scale of the observed cosmological
constant, $v$. For instance, some typical six-dimensional
interactions which could be generated in this way have the
schematic form
\beq
    S_{\rm eff} = \int d^6x \sqrt{-g_6} \; \left[ c_0 M^6 + c_1 M^4
    R + c_2 M^2 R^2 + c_3 \log(M/\mu) R^3 + \cdots \right] \,,
\eeq
where $R$ is the 6D curvature scalar and the arbitrary constants,
$c_i$, are dimensionless and are taken to be $O(1)$. $\mu$ here
denotes an arbitrary scale whose value is not important for the
present purposes. The ellipses describe terms involving the other
fields of 6D supergravity which involve the same number of
derivatives, as well as other curvature invariants {\it etc.}.

Evaluating this action at the SLED vacuum configuration implies $R
\sim 1/r^2$, and --- keeping in mind the volume of the internal
two dimensions is order $r^2$ --- we see that $S_{\rm eff}$
generates the following ultraviolet-sensitive terms in the
effective 4D scalar potential for $r$:
\beq
    \delta V_{\rm eff} = c_0 M^6 r^2 + c_1 M^4 + c_2 M^2/r^2 + c_3
    \log(M/\mu)/r^4 + \cdots \, .
\eeq
Clearly it is the terms in $S_{\rm eff}$ which are proportional to
positive powers of $M$ which are dangerous, and whose absence in
any explicit quantum calculation (like those of \cite{ablm}) must
be explained.

It is 6D supersymmetry (and general covariance) which disposes of
the terms of order $M^6$ and $M^4$ in $S_{\rm eff}$. The terms of
order $M^6$ are excluded because supersymmetry forbids a bare
cosmological constant in six dimensions, and so enforces $c_0 =
0$. A unique set of terms of order $M^4$ are allowed by 6D
supersymmetry, and this is precisely the classical supergravity
action with which we start. A quantum contribution to $c_1$ can be
regarded as a renormalization of the classical action, and so does
not change the above arguments. It does not because the precise
value of the classical couplings (like the 6D Newton constant) is
not important for the cancellation of brane tension and bulk
curvature.

The potentially dangerous terms in $S_{\rm eff}$ are those which
involve squares of the curvature together with their partners
under supersymmetry \cite{branesphere}. Furthermore, explicit
one-loop field-theoretic calculations \cite{Doug} show that the
coefficients, $c_2$, of these terms are not generically zero in
specific 6D supersymmetric field theories. In order to see how
large loop contributions to $c_2$ can be, it is useful to borrow
the counting of coupling constants which would follow if the 6D
theory is regarded as the low-energy limit of string
theory.\footnote{It is natural to think in terms of string theory
to describe the ultraviolet behaviour of this model, since we are
interested in massive states whose scale, $M$, is comparable with
the scale of gravitational physics in the extra dimensions.} In
this case it is the expectation of the 6D dilaton itself which
tracks loops, and the rest of this section argues that because of
the small size predicted for the dilaton field it is only the
one-loop contribution to the curvature-squared contributions which
need vanish in order to keep the effective 4D cosmological
constant small enough.

To see how this works it is useful to make the 6D dilaton explicit
in the above estimates, since it is the {\it v.e.v.} of the
dilaton which plays the role of the bulk coupling. This is most
easily performed in the string frame --- for which the classical
action has the form
\beq \label{classact}
    S_{\rm cl} = c_1 \int d^6x \; \sqrt{-\hat{g}} \; e^{-2\phi}
    \left[\hat{R} + \cdots \right] \,.
\eeq
In this frame an $\ell$-loop correction is accompanied by a power
$e^{2 (\ell - 1) \phi}$ (showing that the classical contribution,
eq.~(\ref{classact}), corresponds to tree level: $\ell =
0$).\footnote{It is the common appearance of $\exp(-2\phi)$ in
front of the Einstein, Maxwell and brane actions which suggests
looking for this 6D model as a low-energy limit of Heterotic
string vacua compactified on $K3$ in the presence of NS5-branes
\cite{FernandoSusha}.}

The size of the cosmological constant, on the other hand, is most
directly seen in the Einstein frame, which is defined by rescaling
the metric so that the coefficient of $R$ is independent of
$\phi$. In 6 dimensions this frame is related to the string frame
by the requirement $\hat{g}_{MN} = e^\phi \, {g}_{MN}$, so that
$\sqrt{-\hat{g}} \, e^{-2\phi} \hat{R} = \sqrt{-g} \, R$, and so
an $\ell$-loop contribution to a curvature-squared term becomes of
order
\beq
    \sqrt{- \hat{g}} \, \hat{R}^2 = \sqrt{-g} \, e^{(2\ell -1)\phi} R^2 \,.
\eeq
This shows that the dangerous contribution to $V_{\rm eff}$ is of
order $c_2 \, M_p^2 \, e^{\phi}/r^2$. To estimate the size of this
contribution requires knowing the size of $\phi$, which the
explicit solutions to the 6D field equations show to be $e^\phi
\sim 1/(Mr)^2$ \cite{ss,rbd,susysphere}. Consequently we see that
the one-loop-generated curvature-squared terms are dangerous ---
being of order $M_p^2/(M^2r^4) = M^2/r^2$, where the relation $M_p
= M^2 r$ is used. By contrast, the contributions at two loops and
beyond are smaller than $O(m_{sb}^4)$ because of the extremely
small size of the bulk coupling, $e^{\phi}$. Within the SLED
proposal the only dangerous bulk-loop contributions to the
cosmological constant arise at one loop, making their vanishing
easier to arrange than usual. Work is underway to determine the
general conditions which the vanishing of this one-loop
contribution requires.

One might worry that there may be phenomenological difficulties
with having bulk couplings this small, but this is not a
difficulty for particles localized on the brane since their
couplings are not proportional to $e^\phi$. (That is, because of
the condition $\lambda = 0$ in eq.~(\ref{lambdaeq}), which was
required by the condition that the brane couplings not break the
classical bulk scale invariance.) This shows that it is precisely
the dilaton/brane coupling which ensures the classical
cancellation between brane tensions and bulk curvatures
\cite{susyads} which also ensures that brane couplings are not
suppressed by the small size of the bulk coupling, $e^\phi$.

\subsection{Consistency With Cosmology and Observations?}

Some of the cosmological implications of the 5 dimensional
`self-tuning' models have been explored \cite{carroll}, inasmuch
as they are believed to change the form of the effective Friedmann
equation as seen by observers during or after Big Bang
Nucleosynthesis \cite{binetruyjim}. One might wonder whether
models of the SLED type might be ruled out on cosmological grounds
even if one were to suspend judgement on microscopic naturalness
issues.\footnote{As mentioned earlier, we do not consider here
potential problems at epochs before nucleosynthesis, such as the
LED and SLED problems of
refs.~\cite{PDG,LEDcosmobounds,susyaddbounds}, since these are not
robust to changes to the details of the early universe cosmology
or of the particle physics of the model.}

There are two aspects to consider here. One might first ask
whether the 6D models predict deviations from the effective 4D
Friedmann equations in the same way as can happen for 5D models.
Although the general statement in 6 or higher dimensions is not
known, we do know that these corrections in 5D are suppressed if
the energies of brane fields are small compared with the brane
tensions, $T \sim M^4$. As such change can be expected to be less
and less important at lower and lower energies, and so are not
likely to be the most dangerous worries at and after
nucleosynthesis, for which typical energy densities are much
smaller than $M^4$.

A more pressing concern at late cosmological times is that the
small nonzero vacuum energy is likely to arise in a
radius-dependent way, and so the vacuum energy will likely really
be a radion potential, $\rho = V(r) \sim v^4 \sim 1/r^4$. Since
$r$ is dynamical, and given that $V$ is minimized for large $r
\sim 1/v$, it is energetically possible for $r$ to be rolling
cosmologically during the present epoch. Since this potentially
leads to a quintessence-like picture \cite{quintessence} for the
dark energy, it is subject to two very strong observational
constraints.

The first bound arises in such a picture because the fluctuations
of $r$ are described by an extremely light scalar, whose mass is
of order $m_r \sim (M_p r^2)^{-1} \sim 10^{-32}$ eV, and which
should couple to matter with gravitational
strength.\footnote{Interestingly, such a small mass for a
gravitationally-coupled scalar can only be stable under
renormalizations in 4 dimensions if the world becomes
extra-dimensional at scales $\Lambda \sim (M_p \, m_r)^{1/2} \sim
v$ \cite{ABRS2}.} If so, why doesn't this kind of model fall afoul
of the strong observational constraints on modifications to
gravity at large distances \cite{largedistgravbounds}?

The second bound comes because the value of $r$ determines the
relative strength of the gravitational and other forces, and so a
conflict with observations can arise if $r$ rolls appreciably
during the cosmological evolution after Big Bang Nucleosynthesis.
This point has been forcefully raised within the context of the
SLED proposal by ref.~\cite{petergianmassimo}.

Both of these constraints were explicitly investigated for a
closely related kind of quintessence model in ref.~\cite{ABRS2}.
There it was found to be possible (but not automatic) to satisfy
both constraints and so to construct realistic cosmologies. It was
found in this reference that these bounds can cause trouble for
generic evolution of the scalar $r$, but that phenomenologically
acceptable cosmologies were also allowed for a reasonably wide
range of initial configurations. In essence the constraints on
changes to $M_p$ over cosmological time scales tends to be evaded
because the $r$ motion is heavily over-damped by cosmic friction.
The constraints on long range forces can be satisfied because the
matter couplings of the light scalar are $r$-dependent, and so can
evolve towards small couplings at the present epoch (which is the
only time where such bounds apply).

Of course, this should only be regarded as an existence proof that
these observationally-successful cosmologies are possible, and a
more detailed studies can and should be done given an explicit
prediction for $V(r)$.

\section{Conclusions and Open Issues}

It is worth closing with a brief summary of the advantages of, and
challenges remaining for, the SLED proposal. On the one hand, it
provides the following uniquely attractive new perspectives on the
cosmological constant problem:
\begin{itemize}
\item By far the strongest motivation for the SLED picture is the
understanding it provides as to why the well-understood particles
of ordinary experience (like the electron) do not contribute
unacceptably to the cosmological constant. There are two
conceptual points which allow it to do so. The first is the
observation that the world is six-dimensional at the energies of
ordinary interest, which implies that the zero-point energy of
particles like the electron are localized sources of curvature in
the extra dimensions rather than directly a contribution to the
cosmological constant. It is the extra-dimensional curvature which
this localized energy produces which cancels the potentially large
contributions of the brane tensions to the effective 4D
cosmological constant.
\item Once the brane fields are integrated out, the same must
then also be done for the bulk modes. This leads to the second
important conceptual point of the SLED proposal, which is the
observation that the bulk gravitational sector is much more
supersymmetric than is the sector involving ordinary matter. In
the SLED picture ordinary matter is {\it not} approximately
supersymmetric, and at the weak scale need not resemble the
supersymmetric standard model (minimal or otherwise).
Nevertheless, this supersymmetry breaking on the brane only causes
a small amount of supersymmetry breaking in the gravitational
sector, $m_{sb} \sim 10^{-3}$ eV, because of the weak strength of
the gravitational couplings. It is this small supersymmetry
breaking in the bulk which explains why the 4D cosmological
constant is nonzero but small, rather than being strictly zero.
\item Although it is motivated by the cosmological constant
problem, the SLED proposal has rich experimental implications for
high-energy accelerators. Since gravitational physics must be at
the TeV scale, there should be observable signals of
extra-dimensional gravity at TeV-scale accelerators like the Large
Hadron Collider just as for the non-supersymmetric LED picture.

This would look very different from what would be expected for
supersymmetric extensions of the Standard Model. In its minimal
form --- what might be called `mSLED' --- the description of
physics between the cosmological constant scale (~$10^{-3}$ eV)
and the 6D gravity scale (~10 TeV) is given by a relatively simple
action $S = S_{b} + S_{B} + S_{int} + S_{ob}$. The first two of
these terms describe the physics on our brane and the physics of
the bulk, and are very precisely known. The minimal choice for the
physics on our brane, $S_{b}$, is the Standard Model itself. The
minimal choice for the physics of the bulk is described by the 6D
Nishino-Sezgin action, $S_{B}$, (or its ungauged limit). $S_{int}$
describes the interactions between the bulk and our brane, and is
less well understood. It need not be supersymmetric, and should be
taken to be the most general possible form consistent with all
unbroken symmetries. $S_{ob}$ describes the unknown physics
residing on any other branes which may be situated about the
compact 2 dimensions.
\item There are also likely to be many experimental implications
for non-accelerator physics. These include deviations from the
Newtonian inverse-square law at sub-millimeter scales,
scalar-tensor-type deviations from gravity over long-distances;
and potential implications for astrophysics, along the lines of
the bounds of ref.~\cite{PDG,LEDastrobounds,susyaddbounds}.
\item Finally, to the extent that the energy cost for changing $r$
is as small as $v^4$, this field is very likely to be
cosmologically rolling today, leading to a quintessence-type
cosmology for the dark energy, possibly along the lines of
ref.~\cite{ABRS2}. In this case there is the attractive
possibility of relating cosmological observations of the dark
energy to the evolution of the shape of the extra dimensions. The
effective 4D field theory would in this case consist of the
low-energy Standard-Model limit coupled to any very-low energy
particles on other branes and to the massless modes of the bulk.
\end{itemize}

Much remains to be done however, including the following open
issues (on which work is currently underway).
\begin{itemize}
\item Above all, the main result which remains to be established
is the verification of the size of the 4D cosmological constant
produced by an explicit quantum integration over the bulk modes.
One way to do so is by finding an explicit derivation of the 6D
Nishino-Sezgin model as a vacuum within string theory, since this
allows an accurate identification of which fields count string
loops.\footnote{See ref.~\cite{stringconnection} for recent
progress along these lines.}
\item It would be instructive to have explicit one-loop calculations
to verify how the renormalization of the topological constraints
is not changed as one integrates out scales between $M$ and $v$.
\item The dynamics of the bulk geometry and its implications
for cosmology are in principle calculable for various 6D vacua
(such as spheres or torii). An understanding of this dynamics is
crucial to understanding why the bulk radius should stabilize at
such large values, $r \sim 1/v$, and why the bulk geometry should
remain unwarped (as it must if the vacuum energy is to be as small
as is observed).
\item The detailed phenomenological implications of the SLED
proposal need to be worked out in detail, for both accelerator and
non-accelerator applications.
\end{itemize}

If these ideas are right, the next decades may bring a fruitful
interplay between microscopic and cosmological observations. We
should be so lucky!

\section*{Acknowledgements}

This work is based on the results of very enjoyable collaborations
with Yashar Aghababaie, Jim Cline, Hassan Firouzjahi, Susha
Parameswaran, Fernando Quevedo, Gianmassimo Tasinato and Ivonne
Zavala. This summary has also benefitted from Fernando Quevedo's
thoughtful criticism. My research has been supported by grants
from NSERC (Canada), FCAR (Quebec) and McGill University.

\end{document}